\documentclass[10pt, a4paper, twocolumn, twoside]{article} 

\usepackage[english]{babel} 

\usepackage{microtype} 

\usepackage{amsmath,amsfonts,amsthm} 

\usepackage[svgnames]{xcolor} 

\usepackage[small, labelfont=bf, up]{caption} 

\usepackage{booktabs} 

\usepackage{lastpage} 

\usepackage{graphicx} 

\usepackage{enumitem} 
\setlist{noitemsep} 

\usepackage{sectsty} 
\allsectionsfont{\usefont{OT1}{phv}{b}{n}} 

\usepackage{geometry} 

\usepackage[T1]{fontenc} 
\usepackage[utf8]{inputenc} 

\usepackage{XCharter} 

\usepackage{hyperref}

\usepackage{journals}

\usepackage{fancyhdr} 
\newcommand{\shorttitle}[1]{\fancyhead[CE]{\textsl{#1}}}
\newcommand{\shortauthors}[1]{\fancyhead[CO]{\textsl{#1}}}

\fancypagestyle{firstpage}{ 
	\fancyhf{}
        \lhead{\vspace*{0.0em} \textit{\footnotesize 21$^{\rm st}$ European Workshop on White
            Dwarfs \\ Castanheira, Campos, Montgomery, eds. \\[-0.2em]
          July 23--27, 2018, Austin, Texas, USA}}
}

\date{}

\newcommand{\authorstyle}[1]{{\large\usefont{OT1}{phv}{b}{n}\color{DarkRed}#1}} 

\newcommand{\institution}[1]{{\footnotesize\usefont{OT1}{phv}{m}{sl}\color{Black}#1}} 

\usepackage{titling} 

\newcommand{\HorRule}{\color{DarkGoldenrod}\rule{\linewidth}{1pt}} 

\pretitle{
	\vspace{-30pt} 
	\HorRule\vspace{10pt} 
	\fontsize{16}{12}\usefont{OT1}{phv}{b}{n}\selectfont 
	\color{DarkRed} 
}

\posttitle{\par\vskip 10pt} 

\preauthor{} 

\postauthor{ 
	\vspace{0pt} 
	{\par\HorRule} 
	\vspace{-10pt} 
}

\newcommand{\newabstract}[1]{
    {\section*{Abstract}
    \bfseries #1}
  }

\usepackage[authoryear]{natbib}
\usepackage{amsmath}
\usepackage{amssymb}
\setcitestyle{aysep={}}

\title{Constraining	Low-Mass White Dwarf Binaries from Ellipsoidal Variations} 

\shorttitle{Constraining White Dwarf Binaries from Ellipsoidal Variations} 

\shortauthors{Bell, Hermes, and Kuszlewicz} 

\author{
        \authorstyle{K.~J.~Bell,$^{1,2,\dagger}$ J.~J.~Hermes,$^{3,4}$ and J.~S.~Kuszlewicz$^{1,2}$}
	\newline\newline 
	$^1$\institution{Max-Planck-Institut f\"{u}r Sonnensystemforschung, Justus-von-Liebig-Weg 3, 37077 G\"{o}ttingen, Germany}\\ 
	$^2$\institution{Department of Physics and Astronomy, Stellar Astrophysics Centre, Aarhus University,
Ny Munkegade 120, 8000 Aarhus C, Denmark}\\ 
	$^3$\institution{Department of Physics and Astronomy, University of North Carolina, Chapel Hill, NC 27599, USA}\\ 
	$^4$\institution{Hubble Fellow}\\ 
	$^\dagger$\institution{bell@mps.mpg.de}\\
      }

\begin{document}

\maketitle 

\thispagestyle{firstpage} 


\newabstract{
  Stars are stretched by tidal interactions in tight binaries, and changes to their projected areas introduce photometric variations twice per orbit. \citet{Hermes2014} utilized measurements of these ellipsoidal variations to constrain the radii of low-mass white dwarfs in eight single-lined spectroscopic binaries. We refine this method here, using Monte Carlo simulations to improve constraints on many orbital and stellar properties of binary systems that exhibit ellipsoidal variations. 
  We analyze the recently discovered tidally distorted white dwarf binary system  SDSS\,J1054$-$2121 in detail, and also 
  revisit the \citet{Hermes2014} sample.
  Disagreements in some cases between the observations, ellipsoidal variation model, and \emph{Gaia} radius constraints suggest that extrinsic errors are present, likely in the surface gravities determined through model atmosphere fits to stellar spectra.

  }


\section{Ellipsoidal Variations of Low-Mass White Dwarfs}

Extremely low-mass (ELM; $\lesssim 0.25\,M_{\odot}$) white dwarfs are created through mass transfer during a common envelope phase of tight binary evolution \citep[e.g.,][]{Nelemans2001}. The universe is not old enough for isolated stars to have formed ELM white dwarfs \citep[e.g.,][]{Kilic2007}. 
Radial velocity variations reveal that most observed ELM white dwarfs are in close binary systems with either white dwarf or neutron star companions \citep[e.g., the ELM Survey,][]{Brown2010}.

\begin{figure*}
  \centerline{\includegraphics[angle=0,width=1.805\columnwidth,trim={0 1cm 0 0},clip]{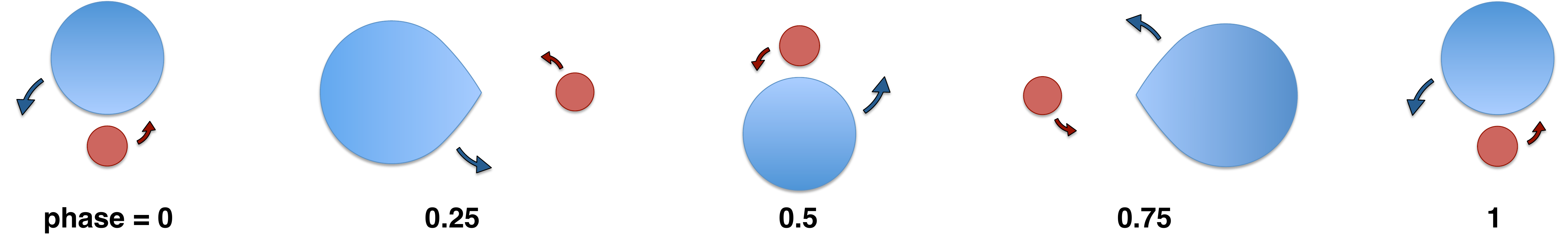}}
  \centerline{\hspace*{-1.66em}\includegraphics[angle=0,width=1.8\columnwidth,trim={0 0.3cm 0 0.3cm},clip]{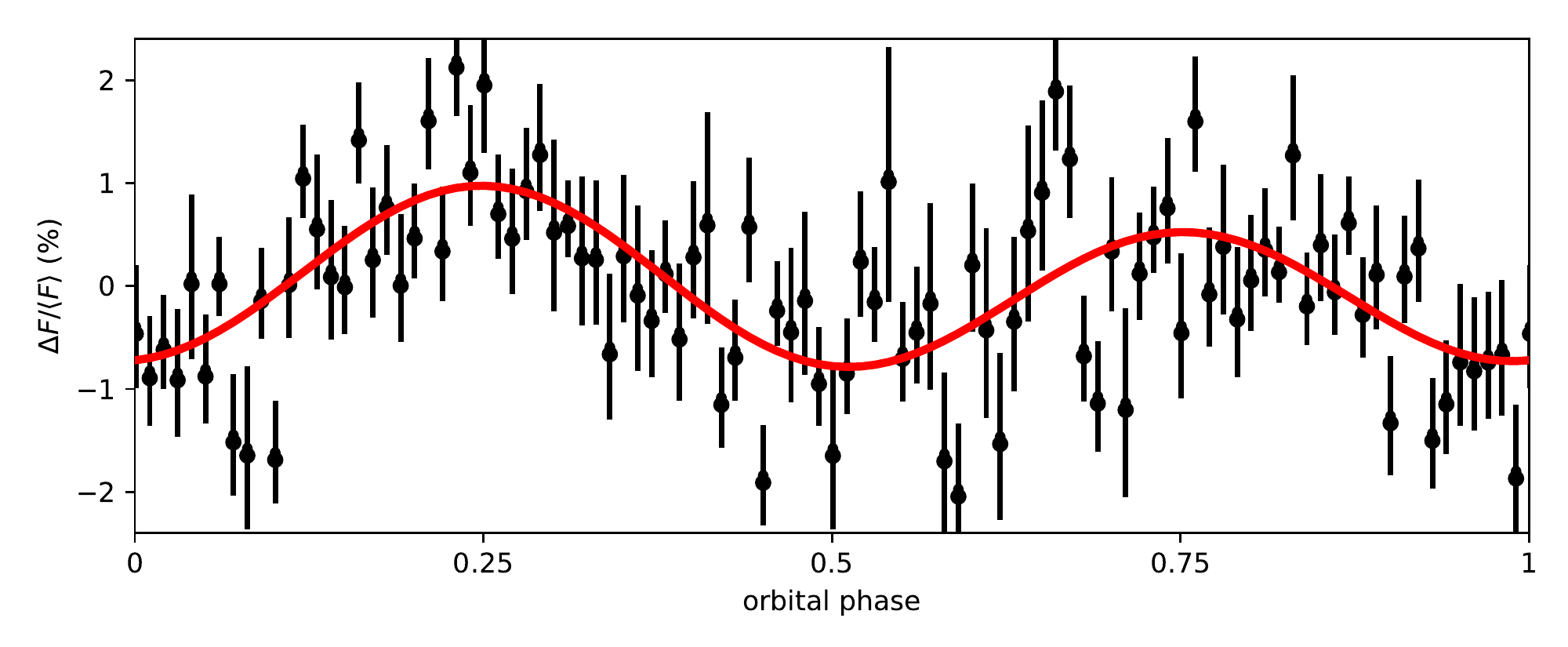}}
  \caption{A cartoon depiction of ellipsoidal variations in a tight binary (not to scale) is displayed above the phase-folded and binned light curve of SDSS\,J1054$-$2121.  The period of the ellipsoidal variation signal is half the orbital period.}
  \label{fig1}
\end{figure*}

Low-mass white dwarfs in tight binaries can be tidally distorted by their
more massive companions. Their projected sizes vary through their
orbits, introducing signatures of ellipsoidal variations to time series
photometric observations \citep[e.g.,][]{Kilic2011,Vennes2011,Brown2011}. 
Ellipsoidal variation signal periods
are half the orbital periods, as demonstrated by the cartoon in Figure~\ref{fig1}.
\citet{Hermes2014} measured amplitudes of these signals from McDonald
Observatory to better constrain the radii of low-mass white dwarf
primaries in eight single-lined spectroscopic binaries. \citet{Bell2017}
detected ellipsoidal variations in another double-degenerate binary,
SDSS\,J1054$-$2121. This phase-folded light curve, averaged within 100 phase bins, is displayed in Figure~\ref{fig1}. The best-fit model to the variations is
overplotted.
The effect of Doppler beaming, which causes the hot primary
to appear brighter when approaching the observer, can also be seen \citep[e.g.,][]{Zucker2007}.

\section{Monte Carlo Rejection \\Sampling}

We refine the Monte Carlo approach of \citet{Hermes2014} to better constrain all parameters of
low-mass white dwarf binaries that exhibit ellipsoidal variations. Here we demonstrate our rejection sampling method, giving specific values for the analysis of SDSS\,J1054$-$2121 \citep[Figure~\ref{fig1};][]{Bell2017}. 

For each star, we draw $10^7$ random deviates from Gaussians representing each of our observed quantities.  Then we sample the probability density functions for the binary parameters of interest by calculating them from each set of deviates.

The ELM Survey \citep[e.g.,][]{Brown2016} provides spectroscopic measurements and uncertainties
for the effective temperatures ($T_\mathrm{eff}$), log surface gravities ($\log{g}$),
orbital periods ($P$), and radial velocity semi-amplitudes ($K_1$) of all of our systems.  We can already use these to constrain other physical properties of these binaries by assuming an isotropic prior on the inclination angle ($i$), as we demonstrate in Section~\ref{sec:withoutEVs}.  In Section~\ref{sec:withEVs}, we include our measurements of the photometric ellipsoidal variation signal amplitudes to improve these constraints considerably.

The approach rests on a few simplifying assumptions: (1) that the rotation periods of the tidally deformed stars $\approx$ the binary periods; (2) that the light curves are records of significant flux from only the primary stars of the single-lined spectroscopic binaries; (3) that the mass of the secondaries are $< 3$\,$M_{\odot}$, corresponding to other white dwarfs or neutron stars; and (4) that there do not exist strong covariances between the measurements of these input quantities.

\begin{figure*}
  \centerline{
  \includegraphics[angle=0,width=0.333\columnwidth,trim={3mm 2mm 2mm 2mm},clip]{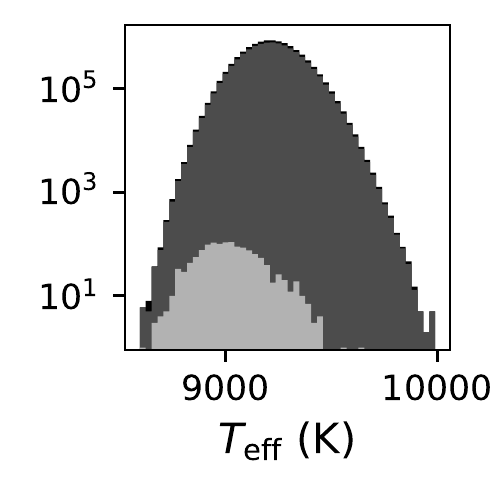}
  \includegraphics[angle=0,width=0.333\columnwidth,trim={3mm 2mm 2mm 2mm},clip]{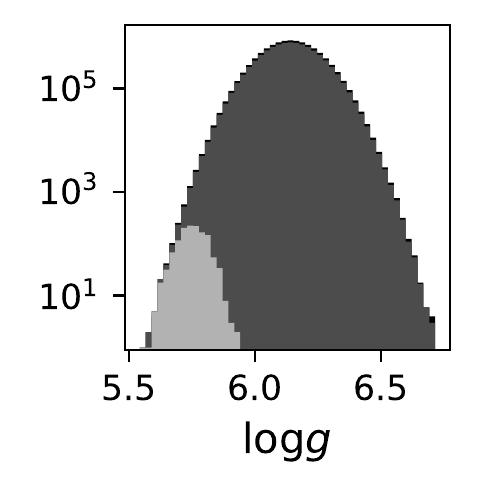}
  \includegraphics[angle=0,width=0.333\columnwidth,trim={3mm 2mm 2mm 2mm},clip]{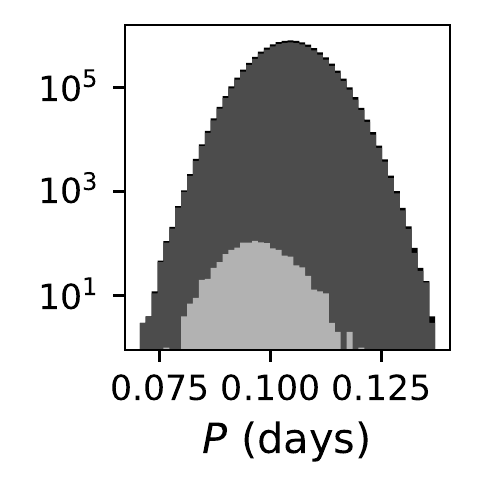}
  \includegraphics[angle=0,width=0.333\columnwidth,trim={3mm 2mm 2mm 2mm},clip]{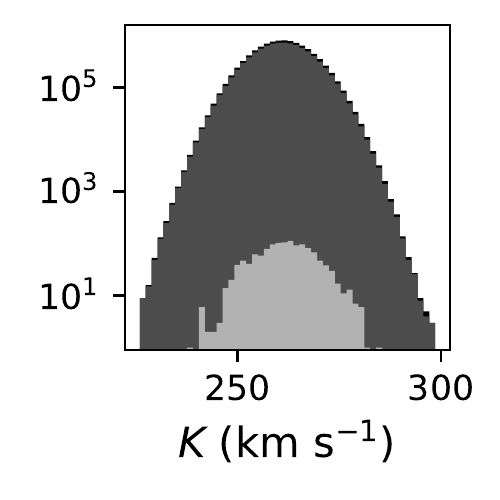}
  \includegraphics[angle=0,width=0.333\columnwidth,trim={3mm 2mm 2mm 2mm},clip]{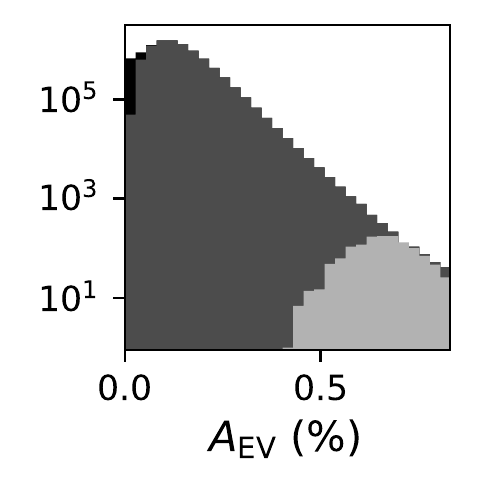}
  }
  \centerline{
  \includegraphics[angle=0,width=0.333\columnwidth,trim={3mm 2mm 2mm 0},clip]{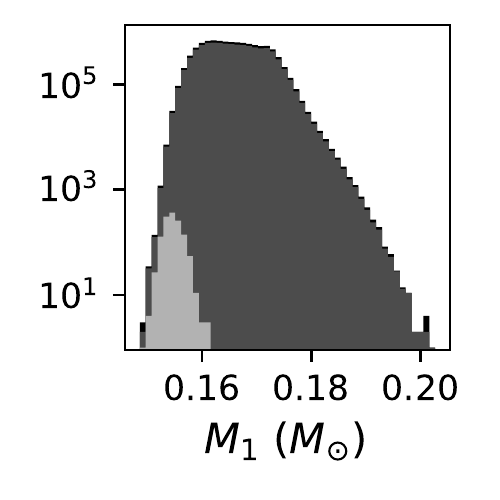}
  \includegraphics[angle=0,width=0.333\columnwidth,trim={3mm 2mm 2mm 0},clip]{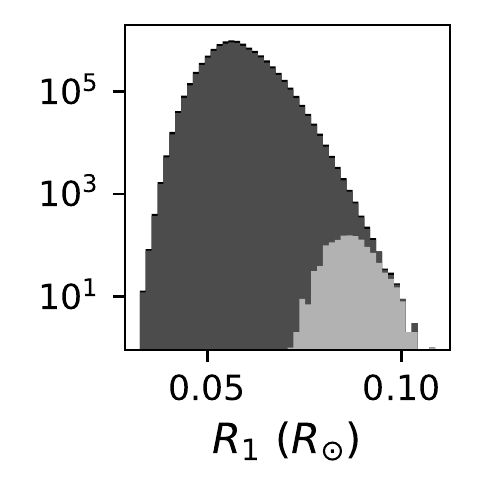}
  \includegraphics[angle=0,width=0.333\columnwidth,trim={3mm 2mm 2mm 0},clip]{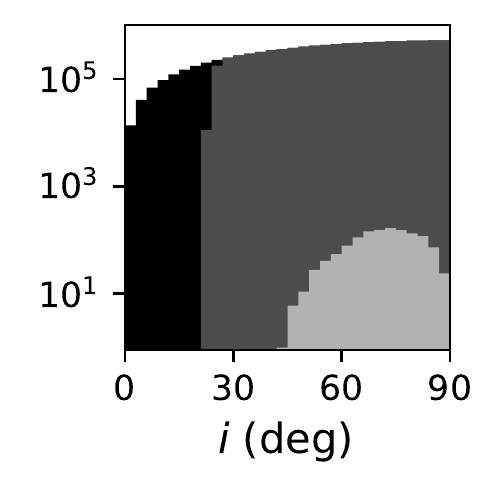}
  \includegraphics[angle=0,width=0.333\columnwidth,trim={3mm 2mm 2mm 0},clip]{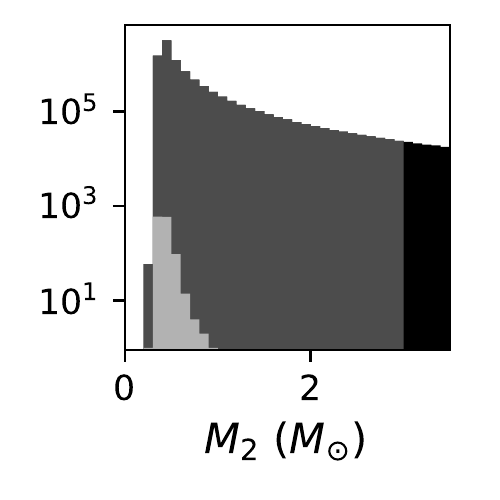}
  \includegraphics[angle=0,width=0.333\columnwidth,trim={3mm 2mm 2mm 0},clip]{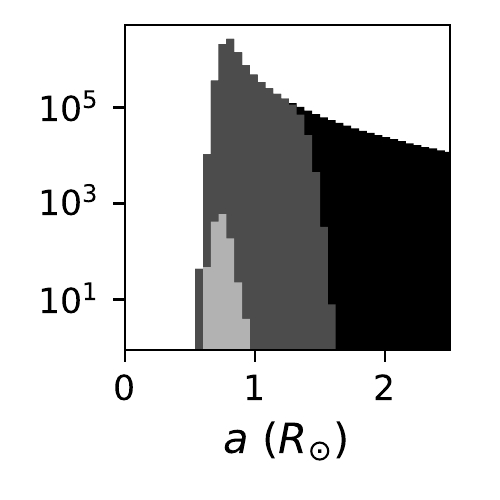}
  \includegraphics[angle=0,width=0.333\columnwidth,trim={3mm 2mm 2mm 0},clip]{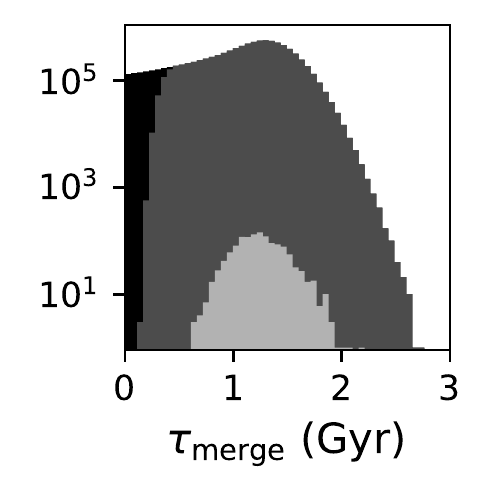}
  }
  \caption{Marginal distributions of Monte Carlo samples for parameters of the binary system SDSS\,J1054$-$2121.  The black histograms represent the original samples (some truncated), the dark gray histograms show samples that satisfy $M_2 < 3$\,$M_{\odot}$ (based on spectroscopy only), and our final constraints that are informed by the measurement of the photometric ellipsoidal variation amplitude, $A_{\rm EV}$, are displayed in light gray. Definitions and summary statistics are in Table~\ref{tab:j1054}.}
  \label{fig:dists}
\end{figure*}

\begin{table*}
	\centering
	\caption{System parameters for SDSS\,J1054$-$2121\label{tab:j1054}.}
	\begin{tabular}{lllcc} 
		\hline
		Parameter Name & Symbol & Units & Spectroscopy Only & With $A_{\rm EV}$\\
		\hline
		orbital period & $P$ & days                     & $0.104\pm0.007$       & $0.096\substack{+0.007 \\ -0.006}$\\
		RV semi-amplitude & $K_1$ & km s$^{-1}$         & $261\pm7$             & $261\pm7$\\
		surface gravity & $\log(g)$ & $g$ in cm s$^{-2}$& $6.14\pm0.11$         & $5.75\pm0.05$\\
		effective temperature & $T_{\rm eff}$ & K       & $9220\pm130$          & $9010\substack{+150 \\ -130}$\\
		ellipsoidal variation amp. & $A_{\rm EV}$ & \%  & $0.13\substack{+0.07\\ -0.05}$   & $0.66\pm0.08$ \\
        primary ELM mass & $M_1$ & $M_{\odot}$          & $0.166\pm0.006$   & $0.1546\substack{+0.0016 \\ -0.0015}$\\
        primary ELM radius & $R_1$ & $R_{\odot}$        & $0.057\substack{+0.007 \\ -0.006}$   & $0.087\pm0.005$\\
        orbital inclination & $i$ & $\deg$              & $63\substack{+19 \\ -23}$   & $72\substack{+9 \\ -10}$\\
        secondary mass & $M_2$ & $M_{\odot}$            & $0.49\substack{+0.50 \\ -0.09}$   & $0.40\substack{+0.06 \\ -0.04}$\\
        star-star separation & $a$ & $R_{\odot}$        & $0.82\substack{+0.16 \\ -0.07}$   & $0.73\substack{+0.05 \\ -0.04}$\\
        merger timescale & $\tau_{\rm merge}$ & Gyr     & $1.2\substack{+0.3 \\ -0.5}$   & $1.2\pm0.2$\\
		\hline
	\end{tabular}
\end{table*}

\subsection{Constraints from Spectroscopy}\label{sec:withoutEVs}

The spectroscopic parameters from the ELM Survey for SDSS\,J1054$-$2121 were provided in \citet{Gianninas2015}: $T_{\rm eff} = 9210\pm140$\,K; $\log{g} = 6.14\pm0.13$; $P_{\rm orb} = 0.104\pm0.007$\,days; $K = 261.1\pm7.1$\,km\,s$^{-1}$. The $\log{g}$ and $T_{\rm eff}$ have been corrected based on 3D convection simulations \citep{Tremblay2015}. 

Without a measured photometric ellipsoidal variation amplitude, we can still progagate these measurements to constrain other properties of these binary systems with a Monte Carlo approach.  We draw $10^7$ random deviates from Gaussians representing each of the observed spectroscopic quantities, as well as random inclination angles from a uniform $\cos{i}$ distribution\footnote{corresponding to isotropy: \url{http://keatonb.github.io/archivers/uniforminclination}}.  For each set of values, we sample the distributions of derived properties as follows:
\begin{itemize}[noitemsep,leftmargin=*,topsep=0pt]
\item We interpolate the linear limb darkening coefficient, $u_1$, for each $\log{g}$ and $T_{\rm eff}$ from \cite{Gianninas2013}. We use the values calculated for the LSST $g$ band as a proxy for {\emph BG40}, which has a similar central wavelength.

\item We calculate temperature-dependent gravity-darkening coefficients, $\tau_1$, following \cite{Morris1985}, using $\beta = 0.25$ \cite[the law of][]{vonZeipel1924} and 5000\,\AA\ as a representative central wavelength of the $BG40$ bandpass.

\item Direct bilinear interpolation of Table 3 from \cite{Althaus2013} gives a mean and standard deviation spread of their evolutionary ELM model masses that could correspond to each $\log{g}$ and $T_{\rm eff}$ deviate pair.  We select a random deviate for $M_1$ from the corresponding Gaussian distribution.

\item The ELM white dwarf radius, $R_1$, follows directly from the definition $g = GM_1/R_1^2$.

\item We calculate the secondary mass, $M_2$, from the measured mass function: 
$$P_{\rm orb}K^3_1/2\pi G = M^3_2\sin^3{i}/(M_1+M_2)^2.$$

\item We calculate the expected photometric semi-amplitude of ellipsoidal variations (in cgs units; \citealt{Morris1993} rearranged by \citealt{Hermes2014}):
$$A_{\rm EV} = \frac{3\pi^2(15+u_1)(1+\tau_1)M_2R^3_1\sin^2{i}}{5P_{\rm orb}^2(3-u_1)GM_1(M_1+M_2)}.$$

\item The orbital separation, $a$, comes from solving Kepler's third law: $a^3 = GP^2(M_1+M_2)/ 4\pi^2$.

\item Finally, we calculate the timescale to a binary merger caused by the release of orbital energy from gravitational radiation using the relation (for mass in $M_{\odot}$, period in hours; \citealt{LL1971})
$$\tau_{\rm merge} = \frac{(M_1+M_2)^{1/3}}{M_1M_2} P^{8/3}\times 10^{-2}{\rm\ Gyr.}$$
\end{itemize}\newpage

Physically, we expect a white dwarf or neutron star secondary with mass $M_2 < 3$\,$M_{\odot}$. We reject solutions that violate this inequality, effectively accepting solutions in proportion to a step function prior on companion mass. The marginal distribution for each parameter from the initial random deviates is displayed in black in Figure~\ref{fig:dists}.  The samples that survive the rejection step based on secondary mass are shown in dark gray.  We summarize our constraints on these parameters in the ``Spectroscopy Only'' column of Table~\ref{tab:j1054} by listing the median values, with uncertainties giving the distances to the 15.9 and 84.1 percentiles.  While this percentile range contains the middle 68.2\% of the samples, many of these distributions are decidedly non-Gaussian, as can be seen in Figure~\ref{fig:dists}.

This Monte Carlo approach allows us to propagate our measurements and priors through combinations of model grids and analytic functions to understand the parameter space of unobserved quantities. An immediately obvious application of this is to identify ELM binary systems that are likely to exhibit ellipsoidal variations for photometric follow-up.

\subsection{Constraints with Ellipsoidal \\ Variations}\label{sec:withEVs}

For systems with measured ellipsoidal variation amplitudes, $A_\mathrm{EV}$, we can further reduce the solution
space to include only those parameter combinations that provide agreement between the
observations and the model. For SDSS\,J1054$-$2121, $A_\mathrm{EV}=0.75\pm0.08$\% \citep{Bell2017}.
We incorporate Gaussian samples for $A_\mathrm{EV}$ into our Monte Carlo
framework and numerically solve for $i$ and $M_2$. We still require $M_2 < 3M_\mathrm{\odot}$. 
For $10^7$ Monte Carlo
samples, we accept solutions in proportion to an isotropic prior on inclination (rejecting
nonphysical $\sin{i} > 1$). 

The light gray histograms in Figure~\ref{fig:dists} demonstrate our improved constraints on the parameters of SDSS\,J1054$-$2121.  The $A_\mathrm{EV}$ measurement most significantly restricts the viable range for orbital inclination. However, we note that only 0.013\% of our samples are not rejected for this particular system, implying that our model and measurements are not in very close agreement. Our final constraints are included in the ``With $A_\mathrm{EV}$'' column of Table~\ref{tab:j1054}.  Our results support that the unseen companion in SDSS\,J1054$-$2121 is likely another white dwarf.

\section{Ensemble Radius--Mass \\ Constraints}

\begin{figure}
  \centerline{\includegraphics[angle=0,width=0.95\columnwidth,trim={4mm 2mm 0 2mm},clip]{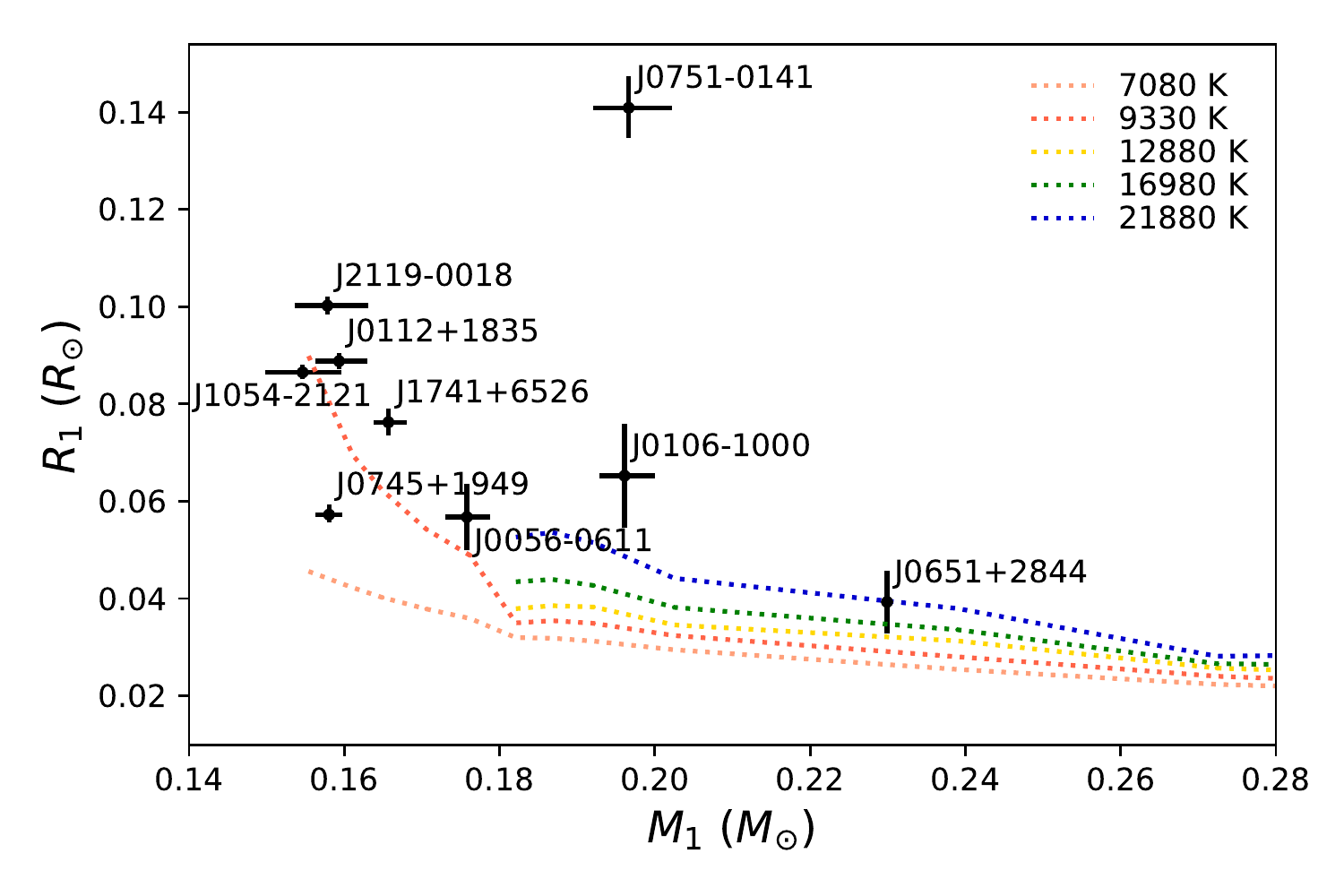}}
  \caption{Radius and mass constraints from Monte Carlo sampling.}
  \label{fig:rm}
\end{figure}

Following the method in Section~\ref{sec:withEVs}, we constrain the parameters of
nine low-mass white dwarf binaries that show ellipsoidal
variations \citep{Hermes2014,Bell2017}.  We plot the mass and radius constraints in Figure~\ref{fig:rm} \citep[mimicking Figure~5 of][]{Hermes2014}.  
We also display the radius--mass
relations at different temperatures
from the evolutionary cooling tracks
of \citet{Althaus2013}. Because these
models are used in our Monte Carlo
calculations, our measurements
follow these tracks by design. The
observations for SDSS\,J0751$-$0141 best fit a pre-white-dwarf model that is still
contracting toward a cooling track.

Our new constraints on $R_1$ and $i$ are compared to the values from 
\citet{Hermes2014} in Table~\ref{tab:allstars}. These agree overall, as they should since they are based on the same measurements. Our new quoted uncertainties are smaller due to a different rejection scheme. The only discrepancy is for SDSS\,J0745+1949, for which only 0.0074\% of our samples (all at the $M_2\approx3M_\odot$ limit) survive rejection, indicating considerable disagreement between the measurements and the model.

\section{Comparison with \emph{Gaia} Radii}

\begin{figure}
  \centerline{\includegraphics[angle=0,width=0.95\columnwidth,trim={4mm 2mm 0 2mm},clip]{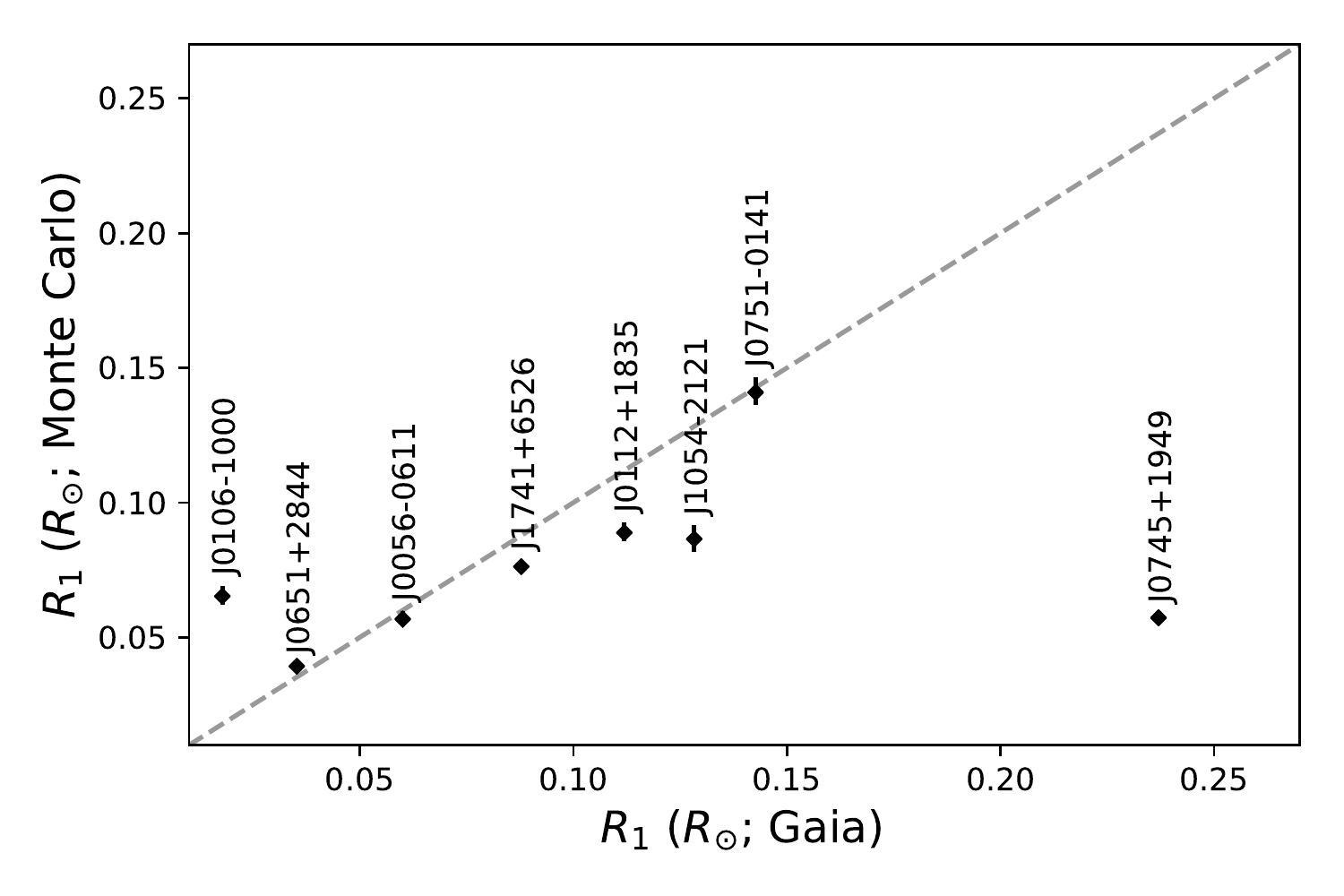}}
  \caption{Comparison of independent radius constraints from Monte Carlo sampling and from \emph{Gaia} astrometry.}
  \label{fig:gaia}
\end{figure}

With the recent availability of \emph{Gaia} astrometry, we can place 
independent constraints on stellar radii based on astrometric distance.
Eight of our stars have positive
parallax measurements in \emph{Gaia} DR2.
We make quick approximations of the distances to these stars by simply inverting Monte Carlo 
samples within the
Gaussian uncertainties of the parallax
measurements.
We then scale the
stellar radii at these distances from
representative DA (hydrogen-atmosphere) white dwarf model
magnitudes\footnote{\url{http://www.astro.umontreal.ca/~bergeron/CoolingModels/}}
 \citep{Holberg2006,Kowalski2006,Tremblay2011} 
 until they match the
observed magnitudes \citep[similar to the solid angle approach of][]{Pelisoli2018}.
We compare the results of our Monte Carlo analysis to the constraints
from \emph{Gaia} in Figure~\ref{fig:gaia}.
The systems that show significant disagreement between these independent radius determinations
are useful for tracing systematic
errors in our measurements, models,
or their interpretation.

\section{Summary and Prospects}

We have demonstrated a work-in-progress Monte Carlo rejection sampling approach for
improving constraints on low-mass white dwarf binary systems that exhibit ellipsoidal variations.
Ellipsoidal variation amplitudes are particularly helpful for constraining the orbital inclination,
and thereby the secondary mass and orbital separation. However, the miniscule fractions of
values that survive rejection sampling for some stars indicate a lack of agreement between
the measurements and the model. These discrepancies provide a tool for identifying
systematic errors in our analysis. The concentration of our final distribution for $\log{g}$ in the far
wing of the prior distribution for SDSS\,J1054$-$2121 in Figure~\ref{fig:dists} suggests that surface gravity may
particularly suffer systematic errors \citep[also suggested in this regime by][]{Brown2017}. This is
corroborated by disagreements with the independent radius constraints from \emph{Gaia} astrometry. Incorporating the
\emph{Gaia} radii into this framework could improve our surface gravity determinations for tidally
distorted white dwarfs and inform our future interpretation of white dwarf spectra.\\

\begin{table*}
	\centering
	\caption{Parameters constrained by ellipsoidal variations for the \citet{Hermes2014} sample.\label{tab:allstars}}
	\begin{tabular}{@{\extracolsep{4pt}}lcccc@{}} 
		\toprule
		 & \multicolumn{2}{c}{\citealt{Hermes2014}} & \multicolumn{2}{c}{This Work}\\
	 \cline{2-3} \cline{4-5}
		\multicolumn{1}{c}{SDSS} & $R_1$ ($R_{\odot}$) & $i$ ($\deg$) & $R_1$ ($R_{\odot}$) & $i$ ($\deg$)\\
		\midrule
		J0056$-$0611& $0.056\pm0.006$ & $50\substack{+22\\-13}$ & $0.057\pm0.002$ & $50\substack{+9\\-7}$ \\
		J0106$-$1000& $0.063\pm0.008$ & $60\substack{+29\\-20}$ & $0.065\substack{+0.004\\ -0.003}$  & $56\substack{+11\\-8}$\\
		J0112+1835  & $0.088\pm0.009$ & $70\substack{+20\\-19}$ & $0.089\substack{+0.004\\ -0.003}$  & $66\substack{+10\\-9}$\\
		J0651+2844  & $0.040\pm0.002$ & $83\substack{+7\\-8}$ & $0.0393\substack{+0.0007\\ -0.0006}$  & $79\substack{+5\\-7}$ \\
		J0745+1949  & $0.176\substack{+0.090\\ -0.025}$ & $63\substack{+27\\-32}$ & $0.0572\pm0.0017$  & $ 10.5\pm0.4$\\
        J0751$-$0141& $0.138\substack{+0.012\\ -0.007}$ & $77\substack{+13\\-17}$ & $0.141\substack{+0.006\\ -0.005}$  & $ 72\pm9$\\
        J1741+6526  & $0.076\pm0.006$ & $78\substack{+12\\-16}$ & $0.0762\substack{+0.0023\\ -0.0019}$ &  $ 75\substack{+7 \\-8 }$\\
        J2119$-$0018& $0.103\pm0.016$ & $75\substack{+15\\-21}$ & $0.100\substack{+0.005\\ -0.004}$ & $ 68\substack{+10 \\-11 }$\\
		\bottomrule
	\end{tabular}
\end{table*}

\noindent \emph{Acknowledgements.}
We gratefully acknowledge support from NSF grant AST-1312983 that funded our data acquisition at McDonald Observatory.
Participation in the 21$^\mathrm{st}$ European Workshop on White Dwarfs was funded by the European Research Council under the European Community's Seventh Framework Programme (FP7/2007-2013) / ERC grant agreement no 338251 (Stellar Ages).  An early version of this work was included in K.J.B.'s PhD Thesis (U.\ Texas). We thank Warren Brown and Mukremin Kilic for comments on the poster, and E.~L.~Robinson for discussions about the method.




\end{document}